\documentclass[10pt,preprint2,authoryear,longnamesfirst]{aastex}
\usepackage{psfig}
\shorttitle{\ion{H}{1} Absorption During 1741$-$038 ESE}
\shortauthors{Lazio et al.}

\newcommand{\kms}{\mbox{km~s${}^{-1}$}}

\newcommand{\cd}{\mbox{cm${}^{-2}$}}

\received{2000 June~30}
\revised{2000 August~10}
\accepted{2000 August~16}
\journalid{546}{2001 January~1} 
\paperid{MS~52199} 
\cpright{PD}{2000}
\begin{document}
\title{The Extreme Scattering Event Toward 1741$-$038: 
	\ion{H}{1} Absorption}

\author{T.~Joseph~W.~Lazio} 
\affil{Naval Research Laboratory, Code~7213, Washington, DC 
        20375-5351, USA}
\email{lazio@rsd.nrl.navy.mil}
 
\author{R.~A.~Gaume} 
\affil{US Naval Observatory, 3450 
	Massachusetts Ave.~NW, Washington, DC  20392-5420, USA}
\email{rgaume@usno.navy.mil} 

\author{M.~J.~Claussen}
\affil{National Radio Astronomy Observatory, P.~O.~Box O, 
	Socorro, NM  87801, USA}
\email{mclausse@nrao.edu}

\author{A.~L.~Fey} 
\affil{US Naval Observatory, 3450 
	Massachusetts Ave.~NW, Washington, DC  20392-5420, USA}
\email{afey@usno.navy.mil}

\author{R.~L.~Fiedler} 
\affil{Naval Research Laboratory, Code~7261, Washington, DC 
        20375-5351, USA}
\email{fielder@sealab.nrl.navy.mil} 

\and 
 
\author{K.~J.~Johnston} 
\affil{US Naval Observatory, 3450 Massachusetts Avenue NW, 
	Washington, DC 20392, USA}
\email{kjj@astro.usno.navy.mil} 

\begin{abstract}
We report multi-epoch VLA \ion{H}{1} absorption observations of the
source \objectname[]{1741$-$038} (\objectname[Ohio]{OT$-068$}) before
and during an extreme scattering event (ESE).  Observations at four
epochs, three during the ESE, were obtained.  We find no changes in
the equivalent width, maximum optical depth, or velocity of maximum
optical depth during the \hbox{ESE}, but we do find a secular trend of
decreasing maximum optical depth between our observations and ones by
other observers a decade prior.  The resulting limit on the \ion{H}{1}
column density change during the ESE for a structure with a spin
temperature~$T_s$ is $6.4 \times 10^{17}\,\cd\,(T_s/10\,\mathrm{K})$.
Tiny-scale atomic structures (TSAS), with $N_H \sim 3 \times
10^{18}$~\cd, are ruled out marginally by this limit, though geometric
arguments may allow this limit to be relaxed.  Galactic halo molecular
clouds, that are opaque in the \ion{H}{1} line, cannot be excluded
from causing the ESE because the observed velocity range covers only
25\% of their allowed velocity range.
\end{abstract}


\section{Introduction}\label{sec:intro}

\begin{figure}
\vspace*{-0.5cm}
\begin{center}
\mbox{\psfig{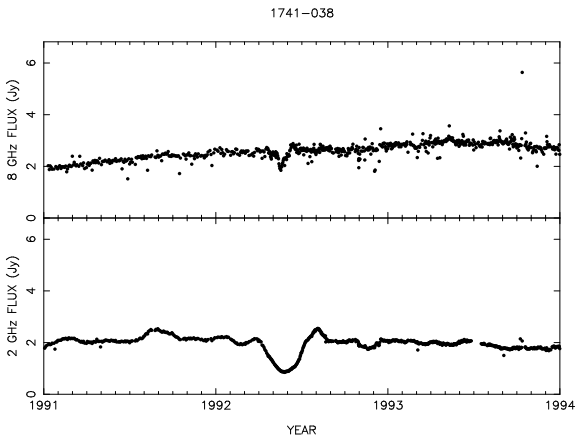}}
%\mbox{\psfig{file=ese.ps,width=0.46\textwidth,angle=-90,silent=}}
\end{center}
\vspace{-0.5cm}
\caption[]{The extreme scattering event toward 1741$-$038.
The upper panel shows the 8.3~GHz light curve, and the lower panel
shows the 2.25~GHz light curve.  Both light curves were obtained with
the Green Bank Interferometer.\label{fig:ese}}
\end{figure}

Extreme scattering events (ESE) are a class of dramatic changes in the
flux density of radio sources \citep{fdjws94}.  They are typically
marked by a decrease ($\gtrsim 50$\%) in the flux density near 1~GHz
for a period of several weeks to months, bracketed by substantial
increases, viz.\ Figure~\ref{fig:ese}.  Because of the simultaneity of
the events at different wavelengths, the time scales of the events,
and light travel time arguments, ESEs are likely due to strong
scattering by ionized structures, possibly in the Galactic
interstellar medium (ISM; \citealp{fdjh87a,rbc87,ww98}).  First
identified in the light curves of extragalactic sources, ESEs have
since been observed during a timing program of the pulsars
\objectname[]{PSR~B1937$+$21} \citep{cblbadd93,lrc98} and
\objectname[]{PSR~J1643$-$1224} \citep{mlc98}.

Modelling of ESE light curves leads to inferred densities $n_e \gtrsim
10^2$~cm${}^{-3}$ within these ionized structures \citep{rbc87,cfl98}.
In turn, these densities imply pressures $nT \sim 10^6$~K~cm${}^{-3}$ or
more, well in excess of the ``average'' interstellar pressure $nT \sim 
3000$~K~cm${}^{-3}$ \citep{kh88}.

A key issue regarding these ionized structures is their relationship
to other phases of the interstellar medium.  Do they represent
(relatively) isolated structures, perhaps in pressure balance with a
lower density, higher temperature ``background'' phase \citep{ccc88}?
or do they reflect a low level of cosmic-ray ionization within an
otherwise neutral structure \citep{h97}? or are they perhaps not
interstellar at all, but due to photoionized molecular clouds in the
Galactic halo \citep{ww98}?

This paper reports the first \ion{H}{1} absorption measurements of a
source (\objectname[]{1741$-$038, OT$-$068}) while it was undergoing
an \hbox{ESE}.  From these observations we can constrain the
connection between the ionized structures responsible for ESEs and any
neutral structures.  In \S\ref{sec:observe} we describe the
observations, in \S\ref{sec:discuss} we discuss the implications of
our observations, and in \S\ref{sec:conclude} we present our
conclusions and suggestions for future work.

\section{Observations and Analysis}\label{sec:observe}

Figure~\ref{fig:ese} shows a portion of the 2.25 and 8.3~GHz light
curve of \objectname[]{1741$-$038} as obtained by the US Navy's
extragalactic source monitoring program at the Green Bank
Interferometer \citep{fiedleretal87b,waltmanetal99}.  Clearly evident
is an approximately 50\% decrease in the source's flux density
at~2.25~GHz and an approximately 25\% decrease at~8.3~GHz.  The
minimum occurred on or near 1992 May~25 (JD~2448768.264), and the
2.25~GHz flux density of the ESE is nearly symmetric about this epoch.
The complete GBI light curve of \objectname[]{1741$-$038}, extending
from~1983 to~1994, has been published previously \citep{cfl98}.

\begin{figure}
\begin{center}
\mbox{\psfig{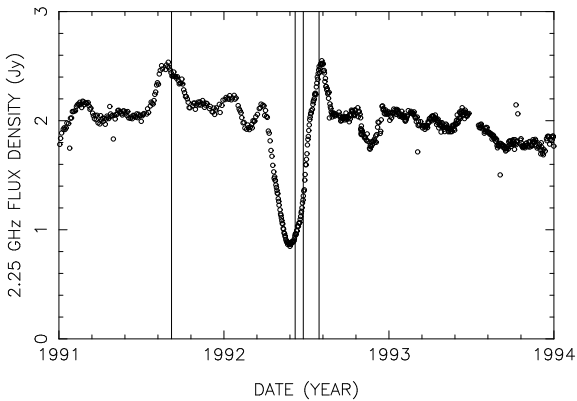}}
%\mbox{\psfig{file=plt_hi.ps,width=0.46\textwidth,angle=-90,silent=}}
\end{center}
\vspace{-0.5cm}
\caption[]{The epochs of \ion{H}{1} absorption during the
ESE of 1741$-$038.  The dots show the 2.2~GHz flux densities measured
by the Green Bank Interferometer.  The vertical lines indicate the
epochs at which \ion{H}{1} absorption observations were
acquired.\label{fig:lightcurve}}
\end{figure}

We used the VLA to measure the \ion{H}{1} absorption toward
\objectname[]{1741$-$038} at four epochs.  Figure~\ref{fig:lightcurve}
shows these four epochs; Table~\ref{tab:log} summarizes the observing
logs for the four epochs.  The first epoch, 1991 September~6, was
prior to the onset of the ESE and shall be used as a control.  The
remaining three epochs occurred during the \hbox{ESE}.  During the
first epoch, full circular polarization was recorded; only right
circular polarization was recorded for the three epochs during the
\hbox{ESE}.  At all four epochs on-line Doppler corrections and
on-line Hanning smoothing were applied.  A bandwidth of~0.78~MHz was
recorded in~256 channels, producing a velocity resolution
of~0.64~\kms.  The observations on 1992 June~6 also included a
comparable amount of observing time with an observing bandwidth
of~1.56~MHz in~256 channels, producing a velocity resolution of
1.29~\kms.  We calibrated these data following the same procedures
described below for the narrower bandwidth observations.  As these
wider bandwidth observations were obtained at only a single epoch, we
will focus largely on the narrower bandwidth observations.  We will
use the wider bandwidth observations only to test for the possibility
of \ion{H}{1} absorption at large velocities.

\begin{deluxetable}{lccccccc}
\tablecaption{VLA Observing Log\label{tab:log}} 
\tabletypesize{\scriptsize}
\tablehead{& \colhead{VLA} & \colhead{Synthesized} &
	& & \colhead{Velocity} & \colhead{On-Source} & \colhead{Flux} \\
        \colhead{Epoch} & \colhead{Configuration} & \colhead{Beam} & 
	\colhead{Polarization} & \colhead{Bandwidth} & \colhead{Resolution} & \colhead{Time} & \colhead{Density} \\
        & & \colhead{(\arcsec)} & 
	& \colhead{(MHz)} & \colhead{(\kms)} & \colhead{(hr)} & \colhead{(Jy)}}

\startdata 
1991 September~6 & A   & 2.5 & R, L & 0.78 & 0.64 & 0.66 & 2.13 \\
1992 June~6\tablenotemark{a} & DnC & 50 & R    & 0.78 & 0.64 & 2.2 & 0.59 \\
1992 June~6\tablenotemark{a} & DnC & 50 & R    & 1.56 & 1.29 & 2.2 & 0.59 \\

1992 June~24     & DnC & 50 & R    & 0.78 & 0.64 & 3.3 & 0.71 \\
1992 July~28     & D   & 50 & R    & 0.78 & 0.64 & 3.5 & 1.92 \\
 
\enddata

\tablenotetext{a}{Observations on 1992 June~6 were obtained
quasi-simultaneously with both 0.78 and 1.56~MHz bandwidths.}
 
\end{deluxetable}

For the three ESE epochs, observing runs consisted of scans on
\objectname[3C]{3C~48} and~\objectname[3C]{3C~286}, for flux density
calibration, and ``on-line'' and ``off-line'' scans of
\objectname[]{1741$-$038}.  On-line scans were centered on an LSR
velocity of~0~\kms\ and included the \ion{H}{1} line.  Off-line scans
were centered at an LSR velocity of~$-360$~\kms; the scans on
\objectname[3C]{3C~48} and~\objectname[3C]{3C~286} were
frequency-shifted by an amount between~$-360$ and~$-442$~\kms.  For
the 1991 September~6 epoch, there were no off-line scans of
\objectname[]{1741$-$038}, the on-line scans of
\objectname[]{1741$-$038} were centered at an LSR velocity of~30~\kms,
only \objectname[3C]{3C~286} was observed for flux density calibration
purposes, and the scans of \objectname[3C]{3C~286} were frequency
shifted to~$-200$~\kms\ and~$+200$~\kms.

\begin{figure}
\vspace*{-0.25cm}
\begin{center}
\mbox{\psfig{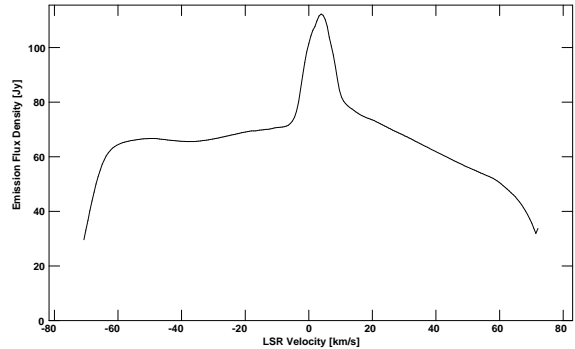}}
%\mbox{\psfig{file=emission.ps,width=0.46\textwidth,angle=-90,silent=}}
\end{center}
\vspace*{-1.25cm}
\caption[]{The \ion{H}{1} emission spectrum as measured by 
a single VLA antenna on 1992 June~24.  No bandpass calibration has
been applied.  Emission spectra from the other two epochs during the
ESE are similar at the 2$\sigma$ level.\label{fig:emission}}
\end{figure}

Figure~\ref{fig:emission} shows the \ion{H}{1} emission spectrum
obtained by a single VLA antenna on 1992 June~24.  The emission
spectra from the other three epochs are similar; in particular, the
spectra at the other two ESE epochs show no deviations above the
$2\sigma$ level within the \ion{H}{1} line.  The shape of the line
on~1991 September~6 is identical, though the maximum amplitude of the
line differs by approximately 10\%.  We would not expect any change
during the ESE because the large primary beam of the VLA antennas
means that they are sensitive to \ion{H}{1} emission on angular scales
much larger than any structures plausibly responsible for the ESE
itself.  The emission spectra we obtain are in generally good
agreement with that found by \cite{dkhv83} using the NRAO 90~m
telescope: The peak emission occurs near~5~\kms, and the emission is
skewed toward positive velocities.  The width of the emission line
from our spectrum and the brightness temperature of the line are
larger than that found by \cite{dkhv83} owing to the smaller diameter
antenna in our case (25~m \textit{vs.} 90~m).

Because we are searching for changes in an absorption line with time,
we discuss in detail the steps we took to calibrate these
observations.  Our primary focus will be on the three observations
with a common bandwidth and velocity resolution during the
\hbox{ESE}.  The calibration of the wider bandwidth observations on
1992 June~6 was identical to the procedure described below.  The
calibration of the first epoch, 1991 September~6, was extremely
similar; we shall only point out the minor ways in which it differed
from the calibration of the other four observations.

The conventional calibration of spectral-line observations involves
determining a bandpass correction.  With such widely distributed
emission as for the \ion{H}{1} line, position-switched observations of
another source cannot be utilized to find the bandpass correction, and
frequency switching requires observations both above and below the
\ion{H}{1} line.  Furthermore, frequency switching can introduce phase
jumps, which would then be propagated into the bandpass correction,
and errors introduced in the frequency switching ultimately limit the
bandpass that can be obtained \citep{wbh97}.  In this case,
frequency-switched observations both above and below the \ion{H}{1}
line were available only on 1991 September~6 and then only for
\objectname[3C]{3C~286}.

More important than the shape of the bandpass during any particular
observation, however, is its stability from one observation to the
next, particularly during the \hbox{ESE}.  In order to assess this
stability, we constructed bandpass corrections for each of the three
ESE epochs using the off-line scans on \objectname[]{1741$-$038}.  For
any given antenna at any epoch, the amplitude of the bandpass
correction was within 8.4\% of unity with a standard deviation of
approximately 5\% and the maximum phase correction was no more than
7\fdg5 with a standard deviation of approximately 5\arcdeg.

We decided ultimately to forego a bandpass correction.  As further
insurance against any bandpass-induced changes, we also excluded the
first and last 16~channels from the analysis.  We also repeated the
calibration described here for the 1992 June~6 observations, including 
various bandpass calibrations determined from the off-line scans of
\objectname[]{1741$-$038}.  We discuss the results in more
quantitative detail below, but the lack of a bandpass calibration
produces no significant change in the absorption spectra we measure.

Amplitude calibration was performed using \objectname[3C]{3C~48}
and~\objectname[3C]{3C~286}.  The frequency offsets between the
on-line scans and the frequencies at which \objectname[3C]{3C~48}
and~\objectname[3C]{3C~286} were observed contribute to less than a 0.1\%
bias in the flux density calibration.  Far more important is the
contribution of the \ion{H}{1} line emission to the system temperature
of the on-line scans.  In order to avoid a bias, we determined the
antenna gain amplitudes for \objectname[]{1741$-$038} using the
off-line scans only, then applied these to both the on- and off-line
scans.  For the 1991 September~6 observations, for which no off-line
scans of \objectname[]{1741$-$038} were available, we used channels
well outside the range over which the \ion{H}{1} emission was seen.
Typical differences between the flux densities on- and off-line were
10--15\%.  As \objectname[]{1741$-$038} is itself a VLA calibrator and
well approximated by a point source over the range of available
baselines, the visibility phases were calibrated using the off-line
scans on \objectname[]{1741$-$038} itself.

The observations during the three ESE epochs were acquired when the
VLA was in its most compact configurations (DnC and D).  In these
configurations the \ion{H}{1} emission may not be entirely resolved
out, particularly on the shortest baselines.  As a result there may be
contamination of the \ion{H}{1} spectra from the changing sky
distribution of the \ion{H}{1} emission moving through the VLA's
sidelobes as the VLA tracks the source \citep{gvs79}.  After amplitude
and phase calibration, we inspected the visibility amplitudes and
phases on the shorter baselines for evidence of \ion{H}{1} line
contamination.  We found that some shorter baselines did show phase
and/or amplitude offsets characteristic of such contamination.  As a
precaution we eliminated baselines shorter than 125~m (600$\lambda$)
from further analysis for all epochs.

Before imaging the data, a single iteration of phase-only
self-calibration was performed on a single continuum channel in the
on-line data.  The starting model was a point source, with the flux
density appropriate for each epoch (Table~\ref{tab:log}).
\objectname[]{1741$-$038} is well approximated by a point source for
all VLA configurations, and we subtracted the continuum in the
visibility domain, utilizing line-free channels of the on-line data to
generate a linear baseline.  We imaged and \textsc{clean}ed each
channel of the data cube separately.  We also produced a continuum
image from the line-free channels of the on-line data (prior to
performing the continuum subtraction).  Absorption spectra were formed
by integrating the continuum-subtracted data cube over a region whose
size was that of the synthesized beam and which was centered on the
position of \objectname[]{1741$-$038}.  These were converted to
opacity spectra using the flux density determined by fitting a
gaussian function to the continuum image.

\begin{figure}
\vspace*{-0.75cm}
\begin{center}
\mbox{\psfig{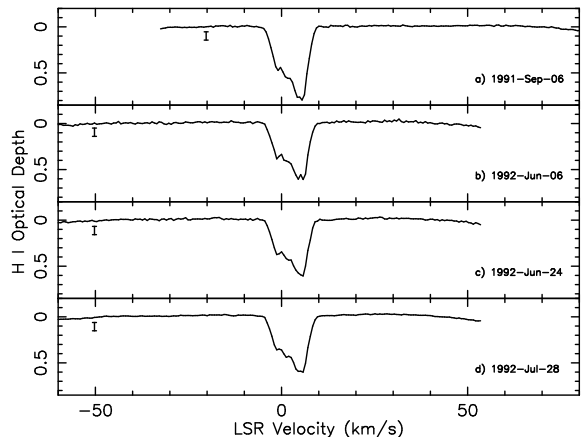}}
%\mbox{\psfig{file=aball.ps,width=0.46\textwidth,angle=-90,silent=}}
\end{center}
\vspace{-0.75cm}
\caption[]{The \ion{H}{1} opacity spectra toward
1741$-$038, see also Figure~\ref{fig:lightcurve}.  All spectra were
obtained by spatially integrating the brightness distribution of
1741$-$038 at every frequency channel over a region comparable to the
half-power width of the synthesized \textsc{clean} beam.  Our angular
resolution is such that 1741$-$038 is unresolved, so we do not show
the images for these epochs.  Shown in the upper left of each figure
is the $\pm 3\sigma$ uncertainty in the optical depth, determined from
a region of the spectrum where there is no absorption.
\textit{(a)}~The epoch 1991 September~6, before the \hbox{ESE}.  This
observation was centered at an LSR velocity of~30~\kms, so the
velocity range is different than for the other three epochs.
\textit{(b)}~The epoch 1992 June~6.  
\textit{(c)}~The epoch 1992 June~24.  
\textit{(d)}~The epoch 1992 July~28.  
\label{fig:absorb}}
\end{figure}

Figure~\ref{fig:absorb} shows the resulting \ion{H}{1} opacity spectra
for the four epochs.  All of the spectra show the presence of a strong
absorption feature near~5~\kms, a slight curvature resulting from the
absence of a bandpass correction, and a typical rms determined outside
the \ion{H}{1} line of $\sigma_\tau \approx 0.012$.  Our absorption
spectra are in good agreement with that found by \cite{dkhv83}.

As a further assurance that our lack of a bandpass calibration has not
introduced significant uncertainties, we repeated the above analysis
using two different bandpass corrections applied to the 1992 June~6
observations.  We applied a bandpass correction determined from the
off-line scans of \objectname[]{1741$-$038} from 1992 June~6 and from
1992 June~24.  In both cases, the spectral baseline after correction
is more nearly constant with velocity.  The rms uncertainty in the
opacity spectrum increases slightly with the application of the 1992
June~6 bandpass correction and little, if at all, with the application
of the 1992 June~24 bandpass correction.  The increase in the rms
uncertainty is approximately 25\% from $\sigma_\tau \simeq 0.012$ to
$\sigma_\tau \simeq 0.015$.  In order to account for the uncertainties
resulting from the lack of a bandpass calibration and from any
possible changes between epochs, we shall use the value $\sigma_\tau =
0.015$ as the value for the typical uncertainty in the opacity
spectrum at all three epochs during the \hbox{ESE}.

\section{Discussion}\label{sec:discuss}

In this section we first assess the extent to which there are any
changes in the absorption spectra.  We then discuss what this implies
about neutral structures along the line of sight to
\objectname[]{1741$-$038} during the \hbox{ESE}.

\subsection{Line Changes}

We shall consider three quantities that might be expected to change
during the ESE---the equivalent width of the \ion{H}{1} line, $W
\equiv \int dv\,\tau(v)$; the maximum absorption and its velocity; and
the shape of the line.  Table~\ref{tab:sum} summarizes these
observable quantities at the four epochs.

\begin{deluxetable}{lccccc}
\tablecaption{Observable Quantities at Each Epoch\label{tab:sum}} 
 
\tablehead{\colhead{Epoch} & \colhead{$\tau_{\mathrm{max}}$} & 
	\colhead{$v$ at $\tau_{\mathrm{max}}$} & \colhead{$W$} & 
	\colhead{$\max(\Delta\tau)$}
	& \colhead{Epoch of $\max(\Delta\tau)$} \\
        &
	& \colhead{(\kms)} & \colhead{(\kms)}}

\startdata
1991 September~6 & 0.79 & 5.5 & 6.27 & \nodata & \nodata \\
1992 June~6  & 0.61 & 5.8 & 4.80 & 0.049 & 1992 June~24 \\
1992 June~24 & 0.61 & 5.8 & 4.77 & 0.049 & 1992 June~6 \\
1992 July~28 & 0.60 & 5.8 & 4.83 & 0.045 & 1992 June~6 \\
 
\enddata 
 
\tablecomments{$\tau_{\mathrm{max}}$ is the maximum optical depth.
$\max(\Delta\tau)$ is the maximum difference in optical depths between
two epochs.  For a given epoch the maximum difference in optical depth
occurs between itself and the epoch listed in Column~6.}
 
\end{deluxetable}

There is no gross change in the absorption line, particularly during
the \hbox{ESE}.  The maximum optical depth for the ESE epochs is $\tau
\simeq 0.61$ and occurs at~5.8~\kms.  The equivalent width is $W
\approx 4.8$~\kms, and from Table~\ref{tab:sum} the maximum difference 
in $W$ between any two epochs is 0.03~\kms.  We estimate the
uncertainty in $W$ as $\sigma_W = \sigma_\tau\Delta v$ where $\Delta
v$ is the velocity extent of the line, $\Delta v \approx 10$~\kms.
Thus, $\sigma_W \approx 0.15$~\kms, and we conclude that there has
been no change in the equivalent width of the line.

One notable difference between the absorption spectra of 1991
September~6 and the ESE epochs is the change in the maximum optical
depth.  The maximum optical depth for all four of our epochs also
differs from previous observations by \cite{dkhv83},
$\tau_{\mathrm{max}} \approx 1$, and \cite{db82}, $\tau_{\mathrm{max}}
\approx 3$.  \cite{db82} observed \objectname[]{1741$-$038} using an
interferometer composed of the NRAO 140~ft and~300~ft telescopes
in~1980 and~1981.  \cite{dkhv83} observed \objectname[]{1741$-$038}
using the phased VLA in~1982.  There is some danger in comparing
observations from different telescopes, and the large optical depth
found by \cite{db82} has a large uncertainty associated with it.
Nonetheless, these changes in the optical depth are indicative of a
decreasing secular trend in the \ion{H}{1} column density by~50\% or
more over the decade between the various observations.

Qualitatively, there do appear to be fine scale changes in the line
shape from epoch to epoch during the \hbox{ESE}.  In particular the
line has two, nearly equally deep minima on 1992 June~6, but only a
single deep minimum in the other two epochs.  We assess the
significance of this difference in the line shape quantitatively by
finding the maximum difference in optical depth between two epochs at
the same velocity (Columns~4 and~5 of Table~\ref{tab:sum}).

The maximum optical depth difference between any two epochs occurs
between 1992 June~6 and June~24 and is $\max(\Delta\tau) = 0.049$.
The maximum difference between 1992 June~6 and July~28 is 0.045.  (The
difference between 1992 June~24 and July~28 is half this value,
reflecting their qualitatively more similar shapes.)  We do not regard
any of these differences as significant.  The rms uncertainty in an
individual opacity spectrum is $\sigma_\tau = 0.015$.  Adding the
uncertainties from two spectra in quadrature, the combined uncertainty
is 0.021, meaning that the maximum deviation is less than 2.3$\sigma$.
We have also examined the difference spectra (i.e., the spectra formed
by taking the difference in the absorption spectra at two epochs) for
any skewness.  If line features either appear or disappear, one would
expect an excess of either positive or negative deviations about the
average.  Conversely, if the deviations are simply the result of
measurement uncertainty, there should be a roughly equal number of
positive and negative deviations about the average.  The latter is the
case for all three difference spectra.

In addition to the line shape not changing by any significant amount,
there is also no evidence for the appearance of additional absorption
components at other velocities.  The only significant absorption
occurs between~$-5$ and~10~\kms\ for all four epochs.  We can also
compare the absorption spectrum from our wider bandwidth (1.56~MHz;
1992 June~6) to those of \cite{dkhv83} as they have a similar velocity
coverage.  Both spectra have a single deep absorption feature, with no
significant absorption outside of it.

\subsection{ESE Structures}\label{sec:structures}

Our limit on changes in the optical depth during the ESE of
$\Delta\tau < 0.05$ implies the neutral column density change during
the ESE must be limited to
\begin{equation}
\Delta N_H < 6.4 \times 10^{18}\,\cd\,\left(\frac{T_s}{100\,\mathrm{K}}\right)
\end{equation}
for the structure(s) responsible for the \hbox{ESE}.  Here $T_s$ is
the spin temperature of the \ion{H}{1} within the structure, and we
have assumed that the structure responsible for this ESE is optically
thin to \ion{H}{1} radiation.  This assumption is justified by the
lack of a large change in the \ion{H}{1} profile.

We consider $T_s \approx 200$~K and $T_s \sim 10$~K as possible values
for $T_s$.  The former value is motivated from observations by
\cite{db82} who find a typical value of $T_s
\sim 200$~\hbox{K} for 14 lines of sight with a Galactic latitude
$10\arcdeg < |b| < 20\arcdeg$ (\objectname[]{1741$-$038} has $b =
13\arcdeg$ and is one of the 14 lines of sight observed).  The latter
value is motivated from models by \cite{h97} and \cite{ww98}, who have
both proposed a population of cold, small-scale structures.

If $T_s \approx 200$~K, then $\Delta N_H < 1.3 \times 10^{19}$~\cd.
This is not a stringent limit.  Models of ESE lenses require electron
column densities~$N_e \sim 10^{16}$~\cd\ \citep{fdjws94,cfl98}.  We
presume that any \ion{H}{1} structure associated with an ESE lens will
have a size scale of order the ESE lens, namely AU scale.  If the
neutral structure is in approximate pressure balance \emph{with the
ESE lens} (which itself may be overpressure with respect to the
ambient medium, \S\ref{sec:intro}), then the neutral structure would
have column densities $N_H \sim 10^{18}$~\cd, well below this
observational limit.

If $T_s \sim 10$~K, then $\Delta N_H < 6.4 \times 10^{17}$~\cd.  This
limit marginally rules out Heiles'~(1997) proposed tiny-scale atomic
structures (TSASs) as being responsible for the \hbox{ESE}.  TSASs are
AU-scale structures distributed throughout the Galactic interstellar
medium with neutral column densities $N_H \sim 3 \times 10^{18}$~\cd
and interior temperatures $T_s \sim 15$~K.  \cite{h97} proposes TSASs
in order to explain observations of small (angular) scale changes in
the \ion{H}{1} opacity, though he does not associate TSASs explicitly
with ESEs.

Obviously, geometrical factors (e.g., a line of sight that does not
cut through the center of a TSAS) could account for the discrepancy.
Other indications of an association between TSASs and ionized
structures are also not clear-cut.  In favor of an association is that
\cite{cfl98} reproduced the ESE light curve of
\objectname[]{1741$-$038} by assuming a gaussian refracting lens
passed in front of the source; they find an electron column density
of~$10^{-4}$~cm${}^{-3}$~pc is required to produce the light curve,
comparable to what the TSASs should have in their interiors due to
photoionization \citep{h97}.  On the other hand a comparable change in
the dispersion measure of PSR~B0823$+$26 occurred \citep{pw91} with no
change in the \ion{H}{1} opacity \citep{fwcm94}.  Though, after the
conclusion of the DM monitoring program, an \ion{H}{1} opacity change
was detected.

Our limit on $\Delta N_H$ appears to rule out Galactic halo molecular
clouds, the AU-scale, \ion{H}{1}-opaque structures that \cite{ww98}
have proposed to explain ESEs.  The photoionized skins of these clouds
would provide the necessary refracting media for ESEs.  We see no
indication of a $\tau > 1$ feature in our spectra (though Walker~2000,
private communication, has since suggested that $\tau \sim 0.1$ might
be more accurate).  \cite{ww98} point out that, because of multiple
imaging, an \ion{H}{1} absorption line during an ESE could saturate
with a non-zero intensity, but \cite{lazioetal00} find no evidence of
multiple imaging in VLBI images of \objectname[]{1741$-$038} acquired
at similar epochs as these \ion{H}{1} absorption measurements.
However, as halo objects, the clouds could have velocities approaching
500~\kms\ (i.e., a velocity range of~1000~\kms).  The velocity range
of our observations, even the wide bandwidth observations of 1992
June~6, is considerably less, being no more than 250~\kms.  Thus, a
significant \ion{H}{1} absorption line could have been present outside
of our velocity range.

All models considered thus far have explained an ESE as being due to
an ionized object, either partially or totally ionized, occulting a
background source.  An alternate possibility is suggested by the
decrease in the \ion{H}{1} optical depth observed between the 1991
September~6 observations and the epochs during the \hbox{ESE}.  If an
ionizing source crossed the line of sight to \objectname[]{1741$-$038}
and ionized some of the hydrogen, the result would be a decrease in
the optical depth of the line.  The difference in equivalent widths
between 1991 September~6 and the epochs during the ESE is 1.47~\kms.
The resulting total change in the column density, integrated over the
line is $N_H = 2.7 \times 10^{19}\,\cd(T_s/10\,\mathrm{K})$.

We assume that any such ionized region would be comparable in extent
to the angular diameter of \objectname[]{1741$-$038}, about 0.5~mas
(Fey, unpublished data).  At a distance of~100~pc, the total quantity
of hydrogen that would have been ionized was $1.2 \times
10^{43}(T_s/10\,\mathrm{K})(D/100\,\mathrm{pc})^2$~atoms.  An estimate
of the ionizing rate required to maintain the ionization for this many
atoms is, following \cite{o89}, at least
$10^{38}\,\mathrm{s}^{-1}(T_s/10\,\mathrm{K})^2(D/100\,\mathrm{pc})$,
assuming that the ionized region was approximately spherical.  An A0
star would suffice as an ionizing source.

\section{Conclusions}\label{sec:conclude}

We have determined the \ion{H}{1} opacity spectra for
\objectname[]{1741$-$038} at four epochs, including three as it
underwent an extreme scattering event (Figure~\ref{fig:lightcurve}).
The absorption spectra from the four epochs are dominated by a strong
absorption feature centered on an LSR velocity of~5~\kms.  The
absorption feature itself is probably a blend of multiple components.
Our absorption spectra are quite similar to a lower-resolution
spectrum obtained by \cite{dkhv83}.

We find a secular trend of decreasing maximum optical depth during the
decade between the observations of \cite{db82}, \cite{dkhv83}, and
those reported here.  In the early 1980s, $\tau \gtrsim 1$ toward
\objectname[]{1741$-$038} while we find $\tau \simeq 0.7$.

We find no evidence for any change in the \ion{H}{1} absorption
feature during the \hbox{ESE}.  Its equivalent width, maximum optical
depth, and velocity at the maximum optical depth are all unchanged for
the three epochs during the \hbox{ESE}.  The maximum optical depth
change between any two epochs during the ESE is $\Delta\tau < 0.05$
($2.3\sigma$) which we do not regard as significant.

The limit on the optical depth change, $\Delta\tau < 0.05$, implies a
limit on the \ion{H}{1} column density of any neutral structure(s)
associated with the ESE lens, $N_H < 6.4 \times
10^{18}\,\cd\,(T_s/100\,\mathrm{K})$, having a spin temperature $T_s =
100$~\hbox{K}.  This limit poses no significant constraint on
structures with $T_s \approx 200$~\hbox{K}.

Some proposed colder structures are allowed.  Tiny-scale
atomic structures \citep{h97}, with $T_s \approx 15$~K and $N_H \sim 3 
\times 10^{18}$~\cd, are marginally ruled out, though geometric
arguments may allow TSASs to be responsible for ESEs and meet this
constraint on the \ion{H}{1} column density.  \cite{ww98} propose
cold, \ion{H}{1}-opaque molecular clouds in the Galactic halo.  Any
such structures within our observed velocity range are clearly
excluded.  However, the observed velocity range covers only 25\% of
the allowed range, so such a cloud could have been responsible for
this ESE without violating our observational constraints.

\acknowledgements
We are grateful to E.~Brinks for many helpful comments about the
analysis of Galactic \ion{H}{1} data.  We are saddened that our thanks
must be posthumous to the referee, R.~Hjellming, for his comments that
helped improve this paper.  We also thank J.~van~Gorkom and
S.~Spangler for motivational commments.  The National Radio Astronomy
Observatory is a facility of the National Science Foundation operated
under cooperative agreement by Associated Universities, Inc.  Basic
research in radio astronomy at the NRL is supported by the Office of
Naval Research.


\begin{thebibliography}{}
\bibitem[\protect\citeauthoryear{Clegg, Chernoff, \& Cordes}{Clegg et
	al.}{1988}]{ccc88} Clegg, A.~W., Chernoff, D.~F., \& Cordes,
	J.~M.  1988, in Radio Wave Scattering in the Interstellar
	Medium, eds.\ J.~M.~Cordes, B.~J.~Rickett, \& D.~C.~Backer
	(New York: American Institute of Physics) p.~174

\bibitem[\protect\citeauthoryear{Clegg, Fey, \& Lazio}{Clegg et
	al.}{1998}]{cfl98} Clegg, A.~W., Fey, A.~L., \& Lazio,
	T.~J.~W.  1998, \apj, 496, 253; astro-ph/9709249

\bibitem[\protect\citeauthoryear{Cognard et al.}{1993}]{cblbadd93} Cognard, I., 
	Bourgois, G., Lestrade, J.-F., Biraud, F.; Aubry, D., Darchy, B., 
	\& Drouhin, J.-P.  1993, \nat, 366, 320

\bibitem[\protect\citeauthoryear{Dickey \& Benson}{1982}]{db82}
	Dickey, J.~M.\ \& Benson, J.~M.  1982, \aj, 87, 278

\bibitem[\protect\citeauthoryear{Dickey et al.}{1983}]{dkhv83} Dickey, J.~M., Kulkarni, S.~R., Heiles,
	C.~E., \& van~Gorkom, J.~H.  1983, \apjs, 53, 59
 
\bibitem[\protect\citeauthoryear{Fielder et al.}{1994}]{fdjws94} Fiedler, R., Dennison, B., 
        Johnston, K.~J., Waltman, E.~B., \& Simon, R.~S.  1994, \apj, 
        430, 581 
 
\bibitem[\protect\citeauthoryear{Fiedler et al.}{1987a}]{fdjh87a} Fiedler, R.~L., Dennison, B.,
	Johnston, K.~J., \& Hewish, A. 1987a, \nat, 326, 675

\bibitem[\protect\citeauthoryear{Fiedler et al.}{1987b}]{fiedleretal87b} Fiedler, R.~L., et al. 
        1987b, \apjs, 65, 319 

\bibitem[\protect\citeauthoryear{Frail et al.}{1994}]{fwcm94} Frail,
	D.~A., Weisberg, J.~M., Cordes, J.~M., \& Mathers, C.  1994,
	436, 144
 
\bibitem[\protect\citeauthoryear{Goss, van~Gorkom, \& Shaver}{Goss et al.}{1979}]{gvs79} Goss, W.~M., van~Gorkom, J.~H., \&
	Shaver, P.~A.  1979, \aap, 73, L17

\bibitem[\protect\citeauthoryear{Heiles}{1997}]{h97} Heiles, C.
	1997, \apj, 481, 193

\bibitem[\protect\citeauthoryear{Kulkarni \& Heiles}{1988}]{kh88} Kulkarni, S.~R.\ \& Heiles, C.
	1988, in Galactic and Extragalactic Radio Astronomy, eds.\
	G.~L.~Verschuur \& K.~I.~Kellermann (Springer-Verlag: Berlin) p.~95

\bibitem[\protect\citeauthoryear{Lazio et al.}{2000a}]{waltmanetal99} Lazio, T.~J.~W., et al. 
        2000a, \apjs, in preparation
 
\bibitem[\protect\citeauthoryear{Lazio et al.}{2000b}]{lazioetal00}
	Lazio, T.~J.~W., et al. 2000b, \apj, 534, 706; astro-ph/9910323

\bibitem[\protect\citeauthoryear{Lestrade, Rickett, \& Cognard}{Lestrade et al.}{1998}]{lrc98} Lestrade, J.-F., Rickett, B.~J., 
        \& Cognard, I.  1998, \aap, 334, 1068 
 
\bibitem[\protect\citeauthoryear{Maitia, Lestrade, \&
	Cognard}{Maitia et al.}{1998}]{mlc98} Maitia, V., Lestrade, J.-F., \&
	Cognard, I.  1998, \apj, submitted

\bibitem[\protect\citeauthoryear{Osterbrock}{1989}]{o89} Osterbrock,
	D.~E.  1989, Astrophysics of Gaseous Nebulae and Active
	Galactic Nuclei (University Science Books: Mill Valley, CA)
 
\bibitem[\protect\citeauthoryear{Phillips \& Wolszczan}{1991}]{pw91}
	Phillips, J.~A.\ \& Wolszczan, A.  1991, \apj, 382, L27

\bibitem[\protect\citeauthoryear{Romani, Blandford, \& Cordes}{Romani et al.}{1987}]{rbc87} Romani, R.~W., Blandford, R.~D., 
        \& Cordes, J.~M. 1987, \nat, 328, 324 
 
\bibitem[\protect\citeauthoryear{Walker \& Wardle}{1998}]{ww98}
	Walker, M.\ \& Wardle, M.  1998, \apj, 498, L125; astro-ph/9802111

\bibitem[\protect\citeauthoryear{Waltman et al.}{1991}]{wfjsfjmm91} Waltman, E.~B., Fiedler, 
        R.~L., Johnston, K.~J., Spencer, J.~H., Florkowski, D.~R., 
        Josties, F.~J., McCarthy, D.~D., Matsakis, D.~N.  1991, \apjs, 
        77, 379

\bibitem[\protect\citeauthoryear{Wilcots, Brinks, \&
	Higdon}{Wilcots et al.}{1997}]{wbh97} Wilcots, E.~M., Brinks, E., \&
	Higdon, J.  1997, A Guide for VLA Spectral Line Observers,
	Edition~9 (NRAO: Socorro, NM); \anchor{http://www.nrao.edu/vla/obstatus/splg1/}{http://www.nrao.edu/vla/obstatus/splg1/}
\end{thebibliography}
\end{document}